\documentclass[twocolumn,tighten]{aastex63}
\usepackage{graphicx}
\usepackage{placeins} 
\usepackage{float}
\usepackage{amsmath,amssymb,bm}
\usepackage[varg]{txfonts}
\usepackage{url}
\usepackage{xcolor} 
\usepackage{natbib}
\usepackage{mathptmx}
\usepackage{flushend}
\usepackage{hyperref,url}
\newcommand{\Bep}{B_\mathrm{ep}}
\newcommand{\Bet}{B_\mathrm{et}}

\received{2021 November 3}
\revised{2022 March 29}
\accepted{2022 April 5}
\shorttitle{Homologous Eruption by Flux Cancellation}
\shortauthors{Hassanin et al.}

\begin{document}

\reportnum{
%$ ~\vspace*{-10cm}
\textit{to appear in ApJ Letters}}

\title{A Model of Homologous Confined and Ejective 
       Eruptions Involving Kink Instability and Flux Cancellation} 

\author[0000-0003-0593-4622]{Alshaimaa Hassanin}
\affiliation{Department of Astronomy, Space Science \& Meteorology, 
	Faculty of Science, University of Cairo, Cairo, Egypt}

\author[0000-0002-5740-8803]{Bernhard Kliem}
\affiliation{Institute of Physics and Astronomy, University of Potsdam, 
14476 Potsdam, Germany}

\author[0000-0003-1624-0802]{Norbert Seehafer}
\affiliation{Institute of Physics and Astronomy, University of Potsdam, 
	14476 Potsdam, Germany}

\author[0000-0003-3843-3242]{Tibor T\"{o}r\"{o}k}
\affiliation{Predictive Science Inc., 9990 Mesa Rim Road, Suite 170, San Diego, CA 92121, USA}

\correspondingauthor{Alshaimaa Hassanin}
\email{shaimaa@sci.cu.edu.eg}

\begin{abstract}

In this study, we model a sequence of a confined and a full eruption, employing the relaxed end state of the confined eruption of a kink-unstable flux rope as the initial condition for the ejective one. The full eruption, a model of a coronal mass ejection, develops as a result of converging motions imposed at the photospheric boundary, which drive flux cancellation. In this process, parts of the positive and negative external flux converge toward the polarity inversion line, reconnect, and cancel each other. Flux of the same amount as the canceled flux transfers to a flux rope, increasing the free magnetic energy of the coronal field. With sustained flux cancellation and the associated progressive weakening of the magnetic tension of the overlying flux, we find that a flux reduction of $\approx\!11\%$ initiates the torus instability of the flux rope, which leads to a full eruption. These results demonstrate that a homologous full eruption, following a confined one, can be driven by flux cancellation. 

\end{abstract}

%% Keywords are replaced by "Unified Astronomy Thesaurus concepts" in AASTeX6.3 which are inserted in the sumbission process
\keywords{Active solar corona, 
          Solar coronal mass ejections, Solar flares, 
          Solar magnetic fields, Solar magnetic reconnection}

\section{Introduction} \label{sec:intro}

A flux rope is a bundle of helically shaped field lines twisting around a common axis. In the solar corona, twisted magnetic fields are also highly sheared, i.e., strongly aligned with a polarity inversion line (PIL) of the photospheric flux. Before a flux rope is formed, the footpoints of magnetic-arcade field lines are always observed to shift along and toward the PIL, due to photospheric motions. A flux rope can then form via driven, slow ``tether-cutting'' reconnection of such a highly sheared field, which is associated with the cancellation of converging flux elements \citep{vanBallegooijen&Martens1989}. The resulting twist is typically around one turn \citep{Mackay&vanBallegooijen2009}, by which a filament can be supported. 

Many filaments lose stability after a sufficient amount of flux shearing and cancellation and experience a confined eruption or evolve into a coronal mass ejection (CME). For example, \citet{Green&al2011} examined a CME on 2007 December~7 from NOAA Active Region (AR) 10977. They found that more than 34\% of the AR flux canceled during the 2.5 days before the eruption, while $\sim\!30\%$ of the AR flux was transformed into the body of the flux rope. Similar results were obtained by \citet{Savcheva&al2012a}. Correspondingly, a category of eruption models assumes that a flux rope is formed before eruption onset and loses equilibrium through an ideal magnetohydrodynamic (MHD) instability (\citealt{vanTend&Kuperus1978, Forbes&Isenberg1991, FanY&Gibson2003, Torok&Kliem2005} (henceforth TK05); \citealt{Kliem&Torok2006, Bobra&al2008}). In contrast, reconnection models \citep{Moore&al2001, Karpen&al2012, JiangC&al2021} propose that a flux rope is formed by the fast ``flare'' reconnection during the eruption, effectively excluding its earlier formation by flux cancellation. 

Eruptions occasionally occur as homologous sequences, i.e., events of similar morphology originating at the same PIL \cite[e.g.,][]{Vemareddy2017, Dhakal&al2020}. Of special interest are series of smaller confined events that can gradually destabilize a filament system and culminate in a CME \citep{Fletcher&Warren2003, ShenYD&al2011, Panesar&al2015, RLiu&al2016, Polito&al2017, WangW&al2019}. They could provide an additional mechanism for the buildup of a flux rope through their flare reconnection, which is topologically identical to tether-cutting reconnection \cite[e.g.,][]{Patsourakos&al2013, Kliem&al2021}. The underlying driver, however, appears to be flux cancellation acting during the sequence of eruptions. This has been seen particularly clearly in miniature versions of this process involving the repeated confined eruption of mini filaments that culminated in a coronal jet \citep{Panesar&al2016, Panesar&al2017}. However, existing models of homologous eruptions rely on either sustained flux emergence \citep{MacTaggart&Hood2009, Chatterjee&FanY2013, Archontis&al2014} or sustained shearing motions \citep{DeVore&Antiochos2008, Soenen&al2009}; neither addresses the role of flux cancellation. Moreover, they produce sequences of confined eruptions or of CMEs only.

Here, we extend these works by presenting the first MHD simulations of a homologous sequence of eruptions that shows a transition from a (not necessarily small) confined eruption to a CME. The sequenceincludes the re-formation of a flux rope after the first, kink-instability-driven eruption, as in \citet{Hassanin&Kliem2016} and \citet{Hassanin&al2016} (henceforth HK16 and HKS16, respectively). The configuration is then driven toward the second, ejective eruption by continuous flux cancellation resulting from imposed photospheric motions converging at the PIL. This forms a new flux rope under the re-formed one, wrapping around its legs. We consider the role of ideal MHD instability, for both flux ropes, vs.\ magnetic reconnection for the onset of the full eruption.

\section{Numerical Model}

A simplified model is adopted for the first eruption, namely that of a pre-existing flux rope experiencing the helical kink instability in the torus-stable regime, which yields a confined eruption (TK05, HK16, HKS16). A more realistic modeling would include the buildup of the initially unstable flux rope by flux cancellation. As this typically yields a kink-stable flux rope, a two-scale photospheric flux distribution would be the most general requirement for the first eruption (triggered by the torus instability) to stay confined \citep{Kliem&al2021}. This is numerically highly demanding and left for a future investigation. 

\begin{figure*} [!t]                                          % Fig. 1
	\centering
	\includegraphics[width=0.8\textwidth,height=0.4\textheight]{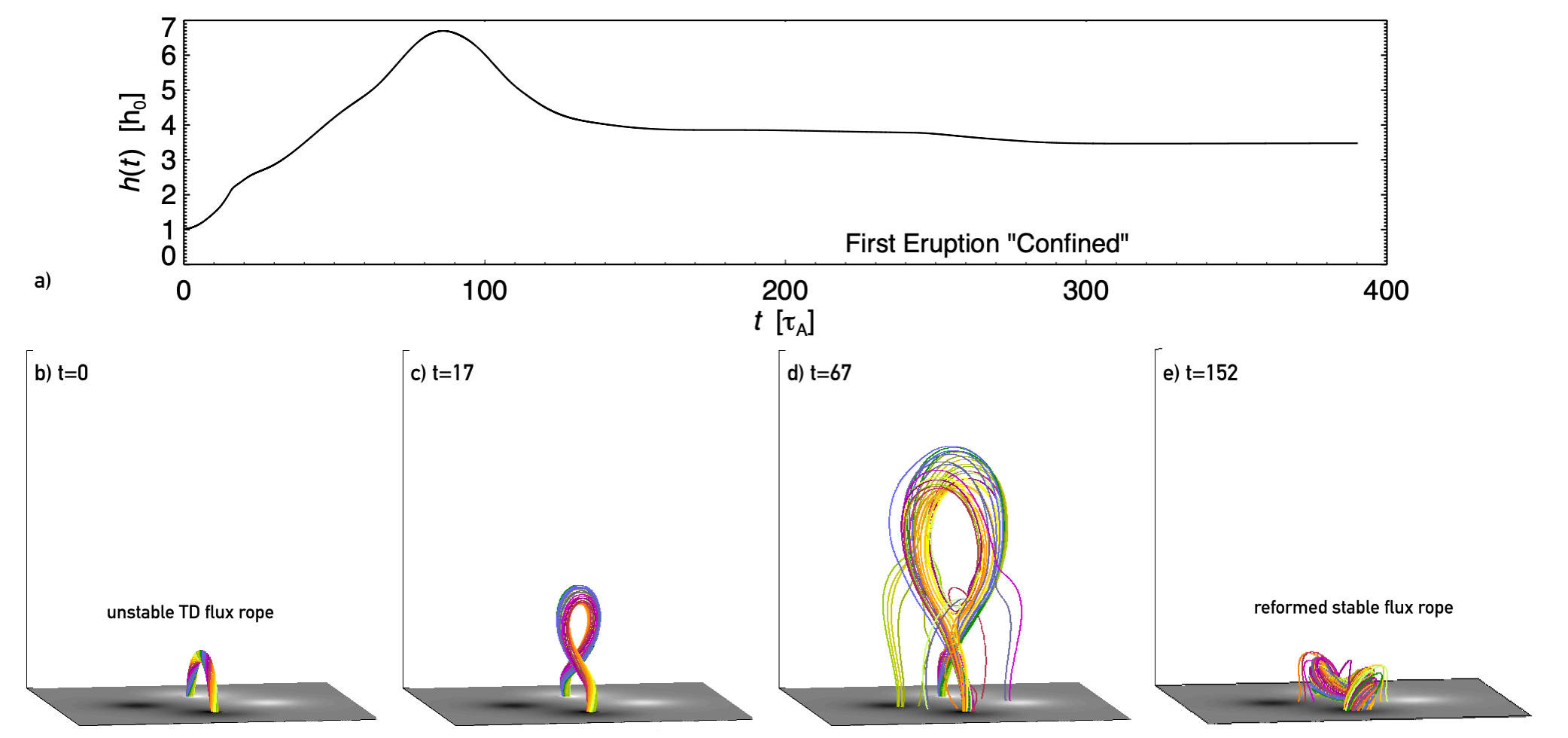} %{confined01.png}
	\caption{Confined eruption.  
                 (a) Rise profile of the fluid element at the apex of the initial flux rope's magnetic axis. 
                 (b)--(e) Main features of the confined eruption.
            } 
	\label{f:time_profile_pert}
\end{figure*}

We start from the same basic simulation setup as in HK16. The force-free magnetic flux rope in the coronal volume, $\{z>0\}$, is a modified \citeauthor{Titov&Demoulin1999} (\citeyear{Titov&Demoulin1999}; henceforth TD99) equilibrium (Figure~\ref{f:time_profile_pert}(b)). The source of the flux-rope field is a toroidal current channel with major and minor radii $R$ and $a$, respectively, placed in the $y$-$z$ plane with the center at $z=-d$ (TD99). The Lorentz self-force of the rope is balanced by the field of two subphotospheric magnetic sources of strength $\pm q$, which are placed at the symmetry axis of the torus at distances $x = \pm L$ from the torus plane and introduce the external poloidal (``strapping'') field, $\Bep$. Here, we replace the line current of TD99, which exerts the external toroidal (``guide'' or ``shear'') field, $\Bet$, with a double-dipole system (TK05). We use the same normalized geometrical parameters as in HK16, except for a smaller distance $L$ of the external polarities (or model ``sunspots'') and a slightly smaller $a$, resulting in $R\approx1.83$, $L=1.75$ and $a\approx0.34$, where lengths are normalized by the initial apex height, $h_0=R-d$. The field strength $B_0$, density $\rho_0$ and corresponding Alfv\'en velocity $V_\mathrm{A0}$ at the apex of the initial flux-rope axis are chosen as further normalization variables. Time is measured by the Alfv\'en time, $\tau_\mathrm{A}=h_0/V_\mathrm{A0}$. The zero-beta, compressible ideal MHD equations, identical to Equations (1)--(4) in HK16, are integrated using a modified two-step Lax-Wendroff scheme \citep{Torok&Kliem2003}. A stretched grid of size $32^3$ is used, with a resolution of 0.02 in the relevant inner part of the box. Closed boundaries are implemented by setting the velocity at the boundaries to zero. 

We deeply relax the configuration resulting from the first eruption and reset the time to zero. We then impose a localized converging-flow pattern at the bottom plane with constant velocity $u_\mathrm{conv}$ and short ramp-up and ramp-down phases, which transports part of the flux from both external polarities toward the PIL. Enhanced numerical magnetic diffusion is introduced at the bottom plane in a strip around the PIL, adjusted such that a strong pileup of the approaching flux elements is avoided.

\section{Results and Analysis}

\subsection{Confined Eruption} 

HK16 and HKS16 demonstrated that the helical kink instability can initiate the eruption of a flux rope and yield a well-fitting model of the confined filament eruption on 2002 May~27 \citep{HJi&al2003, Alexander&al2006}. Here we use the same representative value for the ratio $\Bet/\Bep\approx1$ at the apex of the rope but a slightly higher initial twist of $\phi=4.5\pi$ to emphasize the confinement by the overlying flux even more. The relatively strong guide field yields an opposing force when displaced by the instability. Jointly with the strapping field, this inhibits a full eruption \citep[TK05;][]{Myers&al2015, Filippov2020}. Figure~\ref{f:time_profile_pert}(a) shows that the rope immediately begins to rise, as in HK16 and HKS16. Subsequently, the instability saturates, and the maximum height is reached at $h_\mathrm{max}=6.7$. During the rise, the rope develops a clear helical (inverse-gamma) shape (Figure~\ref{f:time_profile_pert}(c)), which shows a strong conversion of twist into writhe of the rope axis \citep{Torok&al2010}, the typical signature of the helical kink instability. Two reconnection processes follow. The first reconnection occurs between the flux rope and the overlying flux, splitting the top part of the rope (Figure~\ref{f:time_profile_pert}(d)). The reconnected field lines shrink toward the surface. The second reconnection proceeds between the legs of the fully split rope, forming a new rope with a significantly weaker, subcritical twist of $\phi\sim2\pi$ in the torus-stable height range (Figure~\ref{f:time_profile_pert}(e); see also Figure~\ref{f:eruption2}(a) below). The configuration is then deeply relaxed, including enhanced diffusion during $t=240\mbox{--}280$, until the residual velocities stay below $10^{-4}$.

\subsection{Buildup of the erupting flux rope}\label{sec:Building up}

\begin{figure}[ht]                                             % Fig. 2
	\centering
	\includegraphics[width=0.47\textwidth]{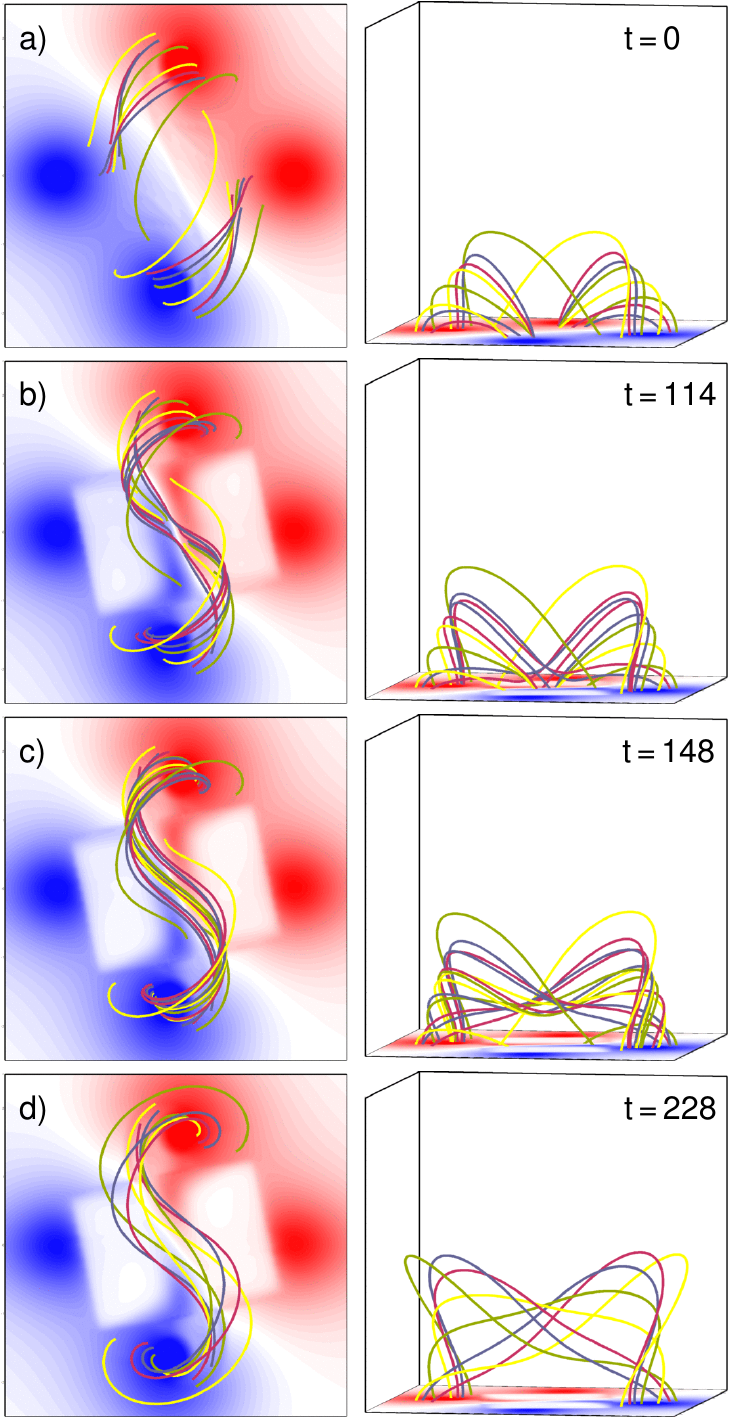} %{fig5_trybk13_v02_version2.png}
	\caption{Transformation of sheared into twisted coronal flux by converging motions and flux cancellation at the photosphere.} 
	\label{f:buildup}
\end{figure}

The imposed converging flows in our model extend between the PIL and the outer area of the model sunspots, mimicking the relevant part of the moat flow on the Sun. We choose the smallest extent in the inflow direction that allows a second eruption. For numerical convenience, we set $u_\mathrm{conv}=0.01$. This is higher than that on the Sun by about an order of magnitude, but sufficiently below the coronal Alfv\'en velocity, ensuring a quasistatic coronal evolution (albeit at velocities typical of the slow-rise phase). As a result, a part of the flux from each sunspot is transported toward the PIL, where it annihilates with the opposite-polarity flux due to magnetic diffusion. This captures the key element of photospheric flux cancellation for the formation and flux feeding of a coronal flux rope---the reconnection of the sheared coronal flux originally rooted in the disappearing photospheric flux \cite[][]{Amari&al2003, Amari&al2011, Aulanier&al2010}. The addition of current-carrying flux energizes the rope and raises its equilibrium height. Simultaneously, the overlying flux is reduced. Both effects weaken the stability of the rope and facilitate its eruption \citep{Mackay&vanBallegooijen2006, Green&al2011, Panesar&al2016, Panesar&al2017, Yardley&al2016}. 

Figure~\ref{f:buildup} shows the transformation of sheared into twisted coronal flux. A part of the flux in the original TD99 rope does not participate in the re-formation during the confined eruption but rather forms sheared loops (Figure~\ref{f:buildup}(a)). The converging motions advect one of their footpoints toward the PIL, thereby strongly increasing their shearand alignment with the PIL (Figure~\ref{f:buildup}(b)). When the photospheric flux they are rooted in cancels (annihilates in our simulation), the loops reconnect across the PIL, detach from the base, and form longer field lines that wrap around the growing flux rope from below, with a total twist of about one turn. The reconnected flux develops a strongly sigmoidal shape because it originates from strongly sheared loops passing over the legs of the re-formed flux rope ($\Bet\sim\Bep$). The triple PIL crossing (Figures~\ref{f:buildup}(b)--(d)) is observed in some soft X-ray sigmoids \cite[e.g.][]{Green&al2011}. 

In the center of the configuration, all reconnected flux runs under the re-formed rope. There is no reconnected flux passing over its apex, although the inflow area covers most of the space between its footprints. We find that the reconnected flux forms a new flux rope, which, although closely wrapped around the legs of the re-formed rope, remains separate in the area of the apex (Figures~\ref{f:eruption2}(b) and (d)). Additionally, all volume currents (except current layers) accumulate in the new rope (Figure~\ref{f:eruption2}(h)). The configuration inherits the nonneutralized current flow from the initial TD99 equilibrium.

\subsection{Ejective Eruption}

\begin{figure*} [!t]   
	\centering                                       % Fig. 3
	\includegraphics[width=\linewidth]{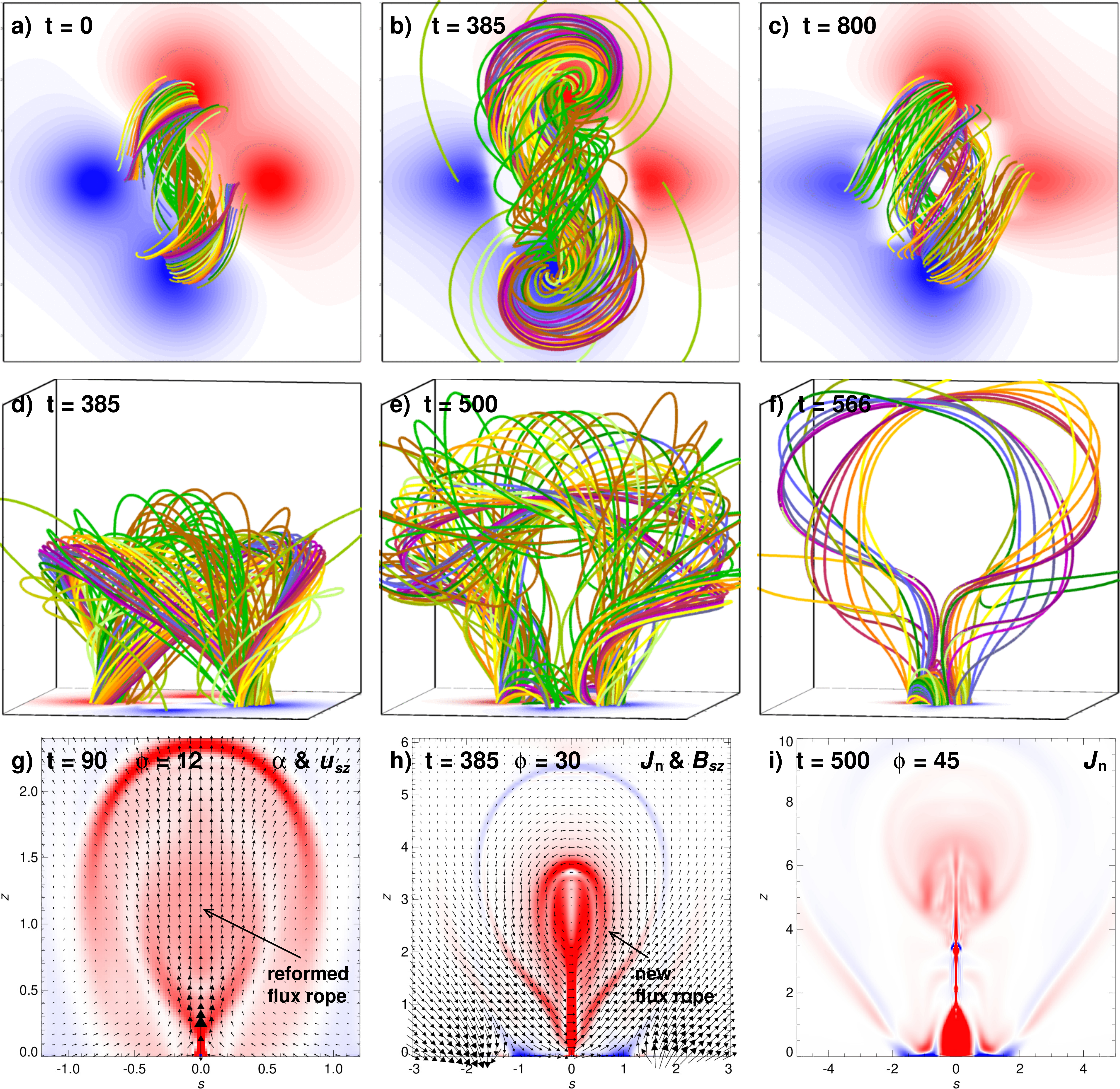} %{Md017_P45_L105_conv_trybk13_Fig2.pdf}
	\caption{Ejective eruption.  
                 (a)--(f) Magnetogram, $B_z(x,y,0,t_i)$, and field lines of the re-formed flux rope (green/brown; drawn from each of the original rope's footprints) 
                 and of the reconnecting flux that forms a new rope (rainbow colors; a subset 
                 is shown in Figure~\ref{f:buildup}). The displayed volume is $7^3$ (a)--(d), $10^3$ (e), and $20^3$ (f). 
                 (g)--(i) Vertical cuts, centered at the $z$-axis and rotated by $\phi$ (in degrees), of the force-free parameter, $\alpha=J_\parallel/B$, and normal (out-of-plane) current-density component, $J_\mathrm{n}$. 
                 Overlaid are in-plane velocity vectors, $u_{sz}$, and field vectors, $B_{sz}$.
                 }
	\label{f:eruption2}
\end{figure*}

\begin{figure*} [!t]                                           % Fig. 4
	\centering
        \includegraphics[width=.8\linewidth]{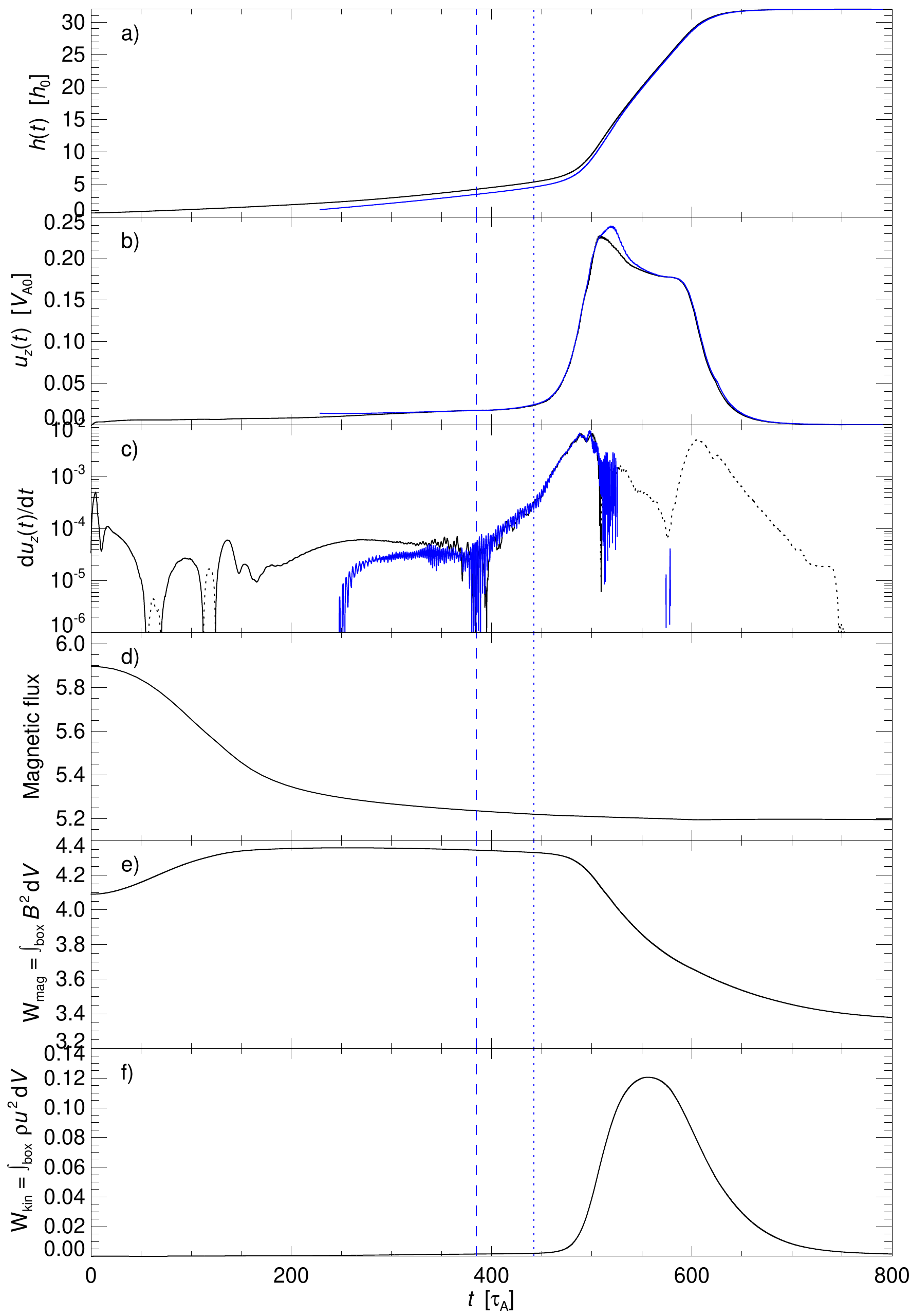} %{f4.eps} %{Md017_P45_L105_conv_new_trybk13_rise+flux+energies.eps}
	\caption{
         (a)--(c) Rise profiles of the fluid elements at the apex points of the re-formed (black) and new (blue) flux ropes after imposing converging photospheric flows (deceleration shown dotted). The latter fluid element is formed at the $z$-axis due to the inflows and reconnection and, therefore, cannot be traced from $t=0$. 
            (d) Magnetic flux, 
            (e) magnetic, and 
            (f) kinetic energies in the box. 
            Vertical dashed (dotted) lines mark the onset of instability (``flare'' reconnection at the $z$-axis).
            } 
	\label{f:time_profile_conv}
\end{figure*}

\begin{figure*}[!t]                                            % Fig. 5
       \centering
       \includegraphics[width=.67\linewidth]{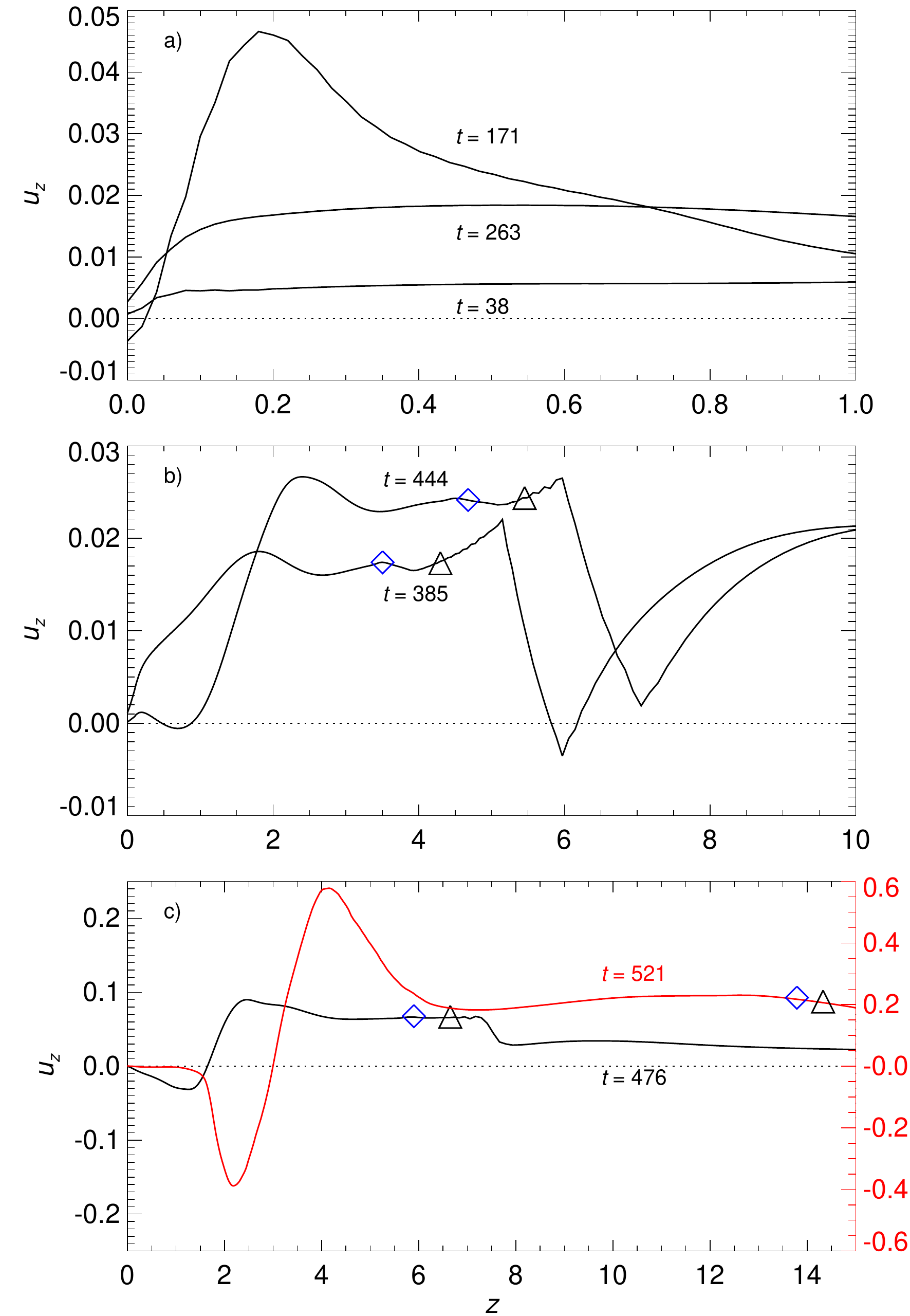} %{f5.eps} %{Md017_P45_L105_conv_trybk13_Fig4.eps}
       \caption{Profiles of $u_z(0,0,z,t_\mathrm{i})$.
                (a) Snapshots before, at peak of, and after ``tether-cutting'' reconnection. 
                (b) Before 
                    and immediately after 
                    the onset 
                    of ``flare'' reconnection flows at the $z$-axis.
                (c) Rapid rise of flare reconnection. 
                    Black and blue symbols mark the flux rope apex positions plotted in Figure~\ref{f:time_profile_conv}(a).
                }
       \label{f:reconn}
\end{figure*}

The photospheric flows drive an upflow above the area of convergence, resulting in the slow inflation of the coronal field \cite[Figure~\ref{f:eruption2}(g);][]{Torok&al2018}. Consequently, the flux ropes show a slow rise (Figures~\ref{f:time_profile_conv}(a--c)), which is also due to their increasing equilibrium height. The rise velocity is of the same magnitude as that of the converging flows, a property verified in the range $u_\mathrm{conv}=0.003\mbox{--}0.03$. The tether-cutting reconnection proceeds in a small vertical current layer low in the box ($z\lesssim0.3$) and decreases strongly with decreasing rate of flux cancellation after $t\approx200$ (Figures~\ref{f:eruption2}(g), \ref{f:time_profile_conv}(d), and \ref{f:reconn}(a)). 

The onset of the second eruption is evident from the onset of exponential acceleration (indicating instability), which is shown by the (similar) rise profiles of both flux ropes. It occurs at $t\approx385$ for the newly formed flux rope and at $t\sim400$ for the re-formed one (Figure~\ref{f:time_profile_conv}(c)). From this delay and the dominance in the current flow, we conclude that the exponential rise of the newly formed flux rope initiates the eruption. 

Up- and downward reconnection outflows reappear at the $z$-axis only at $t=442$ (Figure~\ref{f:reconn}(b)). We cannot exclude an earlier onset away from the $z$-axis in the vertical current sheet that steepens above the PIL at $z\gtrsim0.3$ and $t\gtrsim280$. This is difficult to diagnose in the presence of the guide field and overall upflow. However, if present, the long delay for the spreading of such reconnection to the center of the eruption calls into question a reconnection-dominated onset, as does the fact that the ropes move away from the upper edge of the reconnection outflow at all times (Figure~\ref{f:reconn}). 

The twist of the new flux rope at $t=385$ is numerically estimated from the field in the apex cross section and the length of the approximately determined axis field line to be $\phi<2.5\pi$. This is an upper limit, because the legs are less twisted than the apex region, and reveals stability with respect to the helical kink mode. The decay index, $n=-d \log B_\mathrm{pot,pol}/d \log z$, is determined from the horizontal poloidal component (perpendicular to the new flux rope at its apex) of the potential field computed from the magnetogram at $t=385$. We find $n(z\!=\!3.5)=2.0$ at the position of the fluid element monitored in Figures~\ref{f:time_profile_conv}(a--c) and $n(z\!=\!2.7)=1.7$ at the (slower rising) geometric center of the new flux rope. This indicates clearly that the new flux rope is torus unstable, with the relatively high values of the decay index being consistent with the stabilizing effect of the relatively strong $\Bet$ and the bald-patch topology under the rope (Figure~\ref{f:eruption2}(h)). The even higher decay index value at the position of the re-formed flux rope is irrelevant because only a weak current flows there. We note that, with a weaker initial shear of the reconnecting flux, the new flux rope would develop a higher twist and linkage with the re-formed rope, which could then play a stronger role in the destabilization. 

As in the standard flare model, the vertical current sheet steepens and stretches upward with the rise of the unstable flux (Figure~\ref{f:eruption2}(i)). The resulting, initially amplifying feedback between torus instability and fast, plasmoid-mediated reconnection raises the growth rate of the rise (Figure~\ref{f:time_profile_conv}(c)) and dominates the eruption during the main rise from $t\sim450$ until the upper boundary is approached after $t\sim600$, i.e., a full eruption occurs. The reconnection involves not only ambient flux but also the legs of the flux ropes, which yields two weakly twisted flux bundles connecting each footprint of the original rope with the vicinity of the conjugate footprint (Figure~\ref{f:eruption2}(c)). These would likely merge, again re-forming a flux rope, if further converging flows were applied. The eruptions in our model are homologous---driven by flux ropes with nearly identical footprints, passing over a very similar PIL.

\subsection{Reduction of the photospheric flux}

Figure~\ref{f:time_profile_conv}(d) shows that the total flux continuously decreases with time. It has been demonstrated through modeling of eruption source regions that an eruption occurs only when the axial flux in the rope surpasses a critical value in the range of $\sim\!10\mbox{--}25\%$ of the AR flux \citep{Bobra&al2008, SuYN&al2011}. Combined observational and modeling studies of flux cancellation leading to an eruption mostly found similar values of $\sim\!10\mbox{--}30\%$, in one case $\sim\!50\%$ \citep{Green&al2011, Savcheva&al2012a, Yardley&al2016}. Numerical investigations of eruption triggering by flux cancellation indicate values of 6--10\% \citep{Amari&al2010, Aulanier&al2010}. These relatively low numerical thresholds may result from the greater coherence of the formed flux rope in the coherent field of the numerical models. The reduction of the photospheric flux in the present study amounts to $\approx\!11\%$ of the initial flux by $t = 385$. On the Sun, flux cancellation is driven by photospheric motions that are independent of the onset of eruption. We also let the imposed motions continue past the onset time, until $t=600$. The flux cancellation eventually ceases when the converged flux is nearly completely annihilated at the PIL. Subsequently, the dynamical evolution also ceases.

\subsection{Storage and release of magnetic energy}

The photospheric flux cancellation gradually builds up flux and magnetic energy in the core of the AR and eventually destabilizes the flux rope. The evolution of the total magnetic and kinetic energies is plotted in Figures~\ref{f:time_profile_conv}(e--f). The quasistatic photospheric motions and associated flux cancellation lead to a gradual increase of the magnetic energy up to $t\approx 250$, while the energy of the potential field decreases and the kinetic energy remains negligible. The free magnetic energy, ($W_\mathrm{mag}-W_\mathrm{pot})/W_\mathrm{pot}$, grows from 28\% at $t=0$ to 56\% at $t=385$. As a result of the eruption, part of the magnetic energy is converted into kinetic energy. The free energy then declines to 31\% at $t=800$. For completeness, we note that the ratio of current-carrying to total magnetic helicity in the box peaks at a vlaue of 0.23 at eruption onset ($t=385$), significantly lower than the critical ratio for the onset of the torus instability of $\simeq0.29\pm0.01$ suggested by \citet{Zuccarello&al2018}.

\section{Conclusions and Discussion}

This paper presents a study of flux rope buildup in a sheared field by the flux cancellation and reconnection enforced by photospheric flows converging at the PIL. In full agreement with observations, theoretical concepts \citep{vanBallegooijen&Martens1989} and previous simulations, the process is demonstrated to build free magnetic energy in the topology of a coronal flux rope up to the onset of eruption, due to ideal MHD instability. The sigmoidal rope is weakly twisted ($\sim\!1$ turn), therefore, the torus instability occurs.

The formation of a coherent, unstable flux rope is facilitated by imposing coherent converging flows that extend along a major part of the PIL under the rope. A lower degree of coherence is likely to raise the required amount of canceled flux from our result of $\approx\!11\%$ of the AR flux to a value closer to observational estimates and will be implemented in future work. 

Similar to previous simulations that included flux cancellation driven by converging flows or diffusive transport of magnetic flux toward the PIL \citep{Amari&al2003, Amari&al2011, Aulanier&al2010}, our investigation also demonstrates that an extended vertical current sheet \citep{Mikic&Linker1994} does not form prior to an eruption; a flux rope forms instead. Alternative eruption models require such a current sheet to form before reconnection can initiate an eruption \citep{JiangC&al2021}. Hence, they work only if the converging transport and cancellation of photospheric flux are excluded. 

A new aspect of our model is the combination with a confined eruption, driven here by a kink-unstable flux rope. Employing the relaxed end state as the initial condition for the main simulation, we present a model for homologous eruptions. For the first time, a series of confined and ejective eruptions, and a homologous eruption driven by flux cancellation, are obtained. Because cancellation also weakens the overlying flux, a series of confined homologous events may present a natural path toward a CME, unless the overlying flux is fully restored by the eruptions (HK16; Figure~\ref{f:eruption2}(c)). Extensions of the model are planned to address these aspects and their parametric dependence through more complex sequences of homologous eruptions, driven entirely by flux cancellation.

\acknowledgments

We gratefully acknowledge helpful comments by Dr. M. del Valle, Prof. R. Santos de Lima, and a very constructive report by the referee. 
We appreciate the support from the DAAD, the DFG, L’Or\'eal-UNESCO For Women in Science Young Talents Program-Egypt, from NASA under Grants 80NSSC17K0016, 80NSSC18K1705, 80NSSC19K0082, 80NSSC19K0860, and 80NSSC20K1274, and from NSF's PREEVENTS program under Grant ICER-1854790. T.T. was additionally supported by NRL Contract N0017319C2003 to Predictive Science Inc., a subcontract of NASA LWS grant 80QTR19T0084.

% \bibliographystyle{aasjournal}
% \bibliography{reform}

\end{document}